\documentclass[twocolumn, pra, showpacs]{revtex4-1}
\usepackage{amssymb}
\usepackage{graphicx}
\usepackage{dcolumn}
\usepackage{bm}
\usepackage{amsmath}
\usepackage[colorlinks,linkcolor=red,citecolor=blue]{hyperref}

\begin{document}

\title{Spin squeezing in a spin-orbit coupled Bose-Einstein condensate}
\author{Li Chen$^{1,2}$}
\author{Yunbo Zhang$^{2,3}$}
\email{ybzhang@sxu.edu.cn}
\author{Han Pu$^{4}$}
\email{hpu@rice.edu}

\affiliation{
$^1${Institute for Advanced Study, Tsinghua University, Beijing 100084, China}\\
$^2${Institute of Theoretical Physics and State Key Laboratory of Quantum Optics and Quantum Optics Devices, Shanxi University, Taiyuan 030006, China}\\
$^3${Key Laboratory of Optical Field Manipulation of Zhejiang Province and Physics Department of Zhejiang Sci-Tech University, Hangzhou 310018, China}\\
$^4${Department of Physics and Astronomy, and Rice Center for Quantum
    Materials, Rice University, Houston, TX 77005, USA}
}

\begin{abstract}
We study the spin squeezing in a spin-1/2 Bose-Einstein condensates (BEC) with Raman induced spin-orbit coupling (SOC). Under the condition of two-photon resonance and weak Raman coupling strength, the system possesses two degenerate ground states, using which we construct an effective two-mode model. The Hamiltonian of the two-mode model takes the form of the one-axis-twisting Hamiltonian which is known to generate spin squeezing. More importantly, we show that the SOC provides a convenient control knob to adjust the spin nonlinearity responsible for spin squeezing. Specifically, the spin nonlinearity strength can be tuned to be comparable to the two-body density-density interaction, hence is much larger than the intrinsic spin-dependent interaction strength in conventional two-component BEC systems such as $^{87}$Rb and $^{23}$Na in the absence of the SOC. We confirm the spin squeezing by carrying out a fully beyond-mean-field numerical calculation using the truncated Wigner method. Additionally, the experimental implementation is also discussed.
\end{abstract}

\maketitle

\section{Introduction}
A squeezed spin state (SSS) refers to a quantum state with redistributed spin fluctuations in the phase space spanned by two non-commutative spin operators, such that the quantum fluctuations of one operator is significantly reduced at the expense of the enhanced quantum fluctuations of the other \cite{Kitagawa1993,Ma2011,Pezze2018}. SSS enables precision spin measurement surpassing the standard quantum limit, and has been suggested to be of wide usage in magnetometers \cite{Wineland, Cronin2009}, atomic clocks \cite{Wineland,Bigelow2001}, as well as gravitational-wave interferometers \cite{Gravity}.

Over the past two decades, spinor Bose-Einstein condensates (BECs), due to their high controllability, have been among the most attractive platforms in the research of spin squeezing \cite{Esteve2008,Riedel2010,Gross2010,Sorensen2001,Hamley2012}. A key factor in generating an SSS is the nonlinear interaction of the collective spin, which establishes the correlations among local spins. In the context of a two-component (i.e., spin-1/2) BEC \cite{Ma2011,Sorensen2001,Kawaguchi2012}, the intrinsic density-density interaction provides such a nonlinear interaction. More specifically, the strength of the effective nonlinear spin interaction, which is responsible for generating spin squeezing, is proportional to $g_{\uparrow \uparrow} + g_{\downarrow \downarrow} - 2 g_{\uparrow \downarrow}$, where $g_{\sigma \sigma'}$ represents the interaction strength between the spin components $\sigma$ and $\sigma'$. Unfortunately, for the two most commonly used bosonic species for BEC experiment, $^{87}$Rb and $^{23}$Na, the intra- and the inter-spin interaction strengths are very close to each other \cite{Kawaguchi2012}. As a result, the nonlinear effective spin interaction is two orders of magnitude less than the total density-density interaction strength, and is thus too weak to generate spin squeezing efficiently.

In the present work, we show that the above-mentioned problem can be circumvented by applying artificial spin-orbit coupling (SOC), which has attracted much attention in cold atom research in recent years \cite{Galitski2013,Goldman2014,Zhai2015,WZhang2018}. SOC can induce novel quantum phases and can provide a powerful control knob in quantum gases by controlling the collective behavior of the spatial and the spin degrees of freedom. In the case of a spin-1/2 BEC, the particle-particle collisions establish correlations in the spatial degree of freedom of the atoms which, through the SOC, will in turn establish spin-spin correlations that give rise to spin squeezing. This qualitative picture remains valid even if the intra- and the inter-spin interaction strengths are equal.

In the following, we will provide a detail analysis to confirm this qualitative picture. We present the Hamiltonian and construct a simple two-mode model in Sec. II. The two-mode model allows us to clearly see the emergence of the effective nonlinear spin interaction which leads to the so-called one-axis-twisting spin nonlinearity in generating spin squeezing, and how the strength of this nonlinear spin interaction depends on system parameters. In Sec. III, we employ the truncated Wigner method under the full Hamiltonian and present the results to show that the full numerical calculation confirms the predictions of the simple two-mode model. Finally we conclude in Sec. IV.

\section{Hamiltonian and the Effective Two-mode Model}

\subsection{Full Hamiltonian}
We consider a two-component BEC whose two internal spin states, labeled as $\uparrow$ and $\downarrow$, are coupled by a pair of Raman beams which induces the SOC \cite{Lin2011}. The Hamiltonian takes the following form (we set $\hbar=1$):
\begin{equation}
	H = H_0 + H_\text{int},\label{A1}
\end{equation}
where
\begin{equation}
	H_0 = \int d\mathbf{r}\,\boldsymbol{\hat{\Psi}}^\dagger\left[\frac{\mathbf{k}^2}{2m} - \frac{k_r k_x \sigma_z}{m} + \frac{\Omega}{2}\sigma_x + \frac{\delta}{2}\sigma_z + V_\text{ext}\right]\boldsymbol{\hat{\Psi}}, \label{A2}
\end{equation}
is the single-particle Hamiltonian with $\boldsymbol{\hat{\Psi}} = \left(\hat{\psi}_\uparrow,\hat{\psi}_\downarrow\right)^T$ the spinor field operator, $m$ the atomic mass, $k_r$ the Raman recoil momentum, $\sigma_{x,y,z}$ the Pauli spin operators, $\Omega$ the Raman coupling strength, $\delta$ the two-photon Raman detuning, and $V_\text{ext}$ the external potential. We have assumed that the Raman recoil momentum is along the $x$-axis, hence the motion along the $y$- and the $z$-axes are decoupled from the Raman transition. This allows us to treat the system as an effectively one-dimensional one and greatly simplifies the computation without losing physical insights. The $H_\text{int}$ in Eq.~(\ref{A1}) is the two-body interaction and takes the form:
\begin{equation}
	H_\text{int} = \frac{1}{2}\sum_{\sigma\sigma'}g_{\sigma\sigma'}\int d^3r \hat{\psi}^\dagger_\sigma \hat{\psi}^\dagger_{\sigma'}\hat{\psi}_{\sigma'}\hat{\psi}_{\sigma},\label{A3}
\end{equation}
where $g_{\sigma\sigma'}>0$ characterizes the interaction strength in different spin channels $\sigma,\sigma'=\{\uparrow,\downarrow\}$. In this work, we only focus on the case of two-photon resonance with $\delta=0$, and assume the intra-spin interaction strengths to be equal, i.e.  $g_{\uparrow\uparrow}=g_{\downarrow\downarrow}=g$. Under such a case, the interaction is SU(2) symmetric if $g_{\uparrow\downarrow}=g$.

\begin{figure}[t]
	\includegraphics[width=0.49\textwidth]{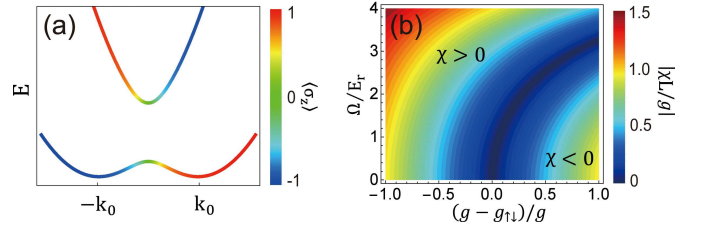}
	\caption{(a) Dispersion of the single-particle Hamiltonian, where the two degenerate ground states $k_x={\pm k_0}$ for an effective spin-1/2 system. (b) Dependence of the two-mode spin nonlinearity strength $\chi$ on $(g-g_{\uparrow\downarrow})$ and $\Omega$.   }
	\label{Fig1}
\end{figure}

For a homogeneous system, i.e. $V_\text{ext}=0$, $k_x$ is a good quantum number, and the single-particle spectrum $E(k_x)$ can be obtained by directly diagonalizing the single-particle Hamiltonian $H_0$ in momentum space. $E(k_x)$ takes the well-known two-band structure \cite{Lin2011}, and is shown in Fig.~\ref{Fig1}(a). The single-particle spectrum has the following features: in the case of $\Omega\in[0,4E_r]$ with $E_r=k_r^2/2m$ being recoil energy, there are two degenerate ground states with ground-state energy $E_0$ and momenta $k_x = \pm k_0$ where $k_0 = k_r \sqrt{1-\Omega^2/16E_r^2}$. Correspondingly, the ground-state wave functions are in the plane-wave forms of
\begin{equation}
	\boldsymbol{\Phi}_{\pm k_0} = \frac{1}{\sqrt{L}}
	\begin{pmatrix}
		\cos \theta \\
		-\sin \theta
	\end{pmatrix}
	e^{i k_0 x}
\text{ and }
 \frac{1}{\sqrt{L}}
	\begin{pmatrix}
		\sin \theta \\
		-\cos \theta
	\end{pmatrix}
	e^{-i k_0 x},\label{A4}
\end{equation}
where $\theta = \arccos(k_0/k_r)/2 \in[0,\pi/4]$ accounts for the spin dressing induced by the Raman coupling, and $L$ is the length of the system. The spin dressing is manifested as the longitudinal polarization $\left\langle \sigma_z\right\rangle = \int dx \boldsymbol{\Phi}^*_{\pm k_0} \sigma_z \boldsymbol{\Phi}_{\pm k_0}  $ deviates from $\pm1$, as is also shown in Fig.~\ref{Fig1}(a). On the other hand, in case of $\Omega>4E_r$, the two local minima merge together at zero momentum $k_x=0$ and the ground state is simply a transversely polarized state $\boldsymbol{\Phi} = L^{-1/2}(1/\sqrt{2},-1/\sqrt{2})^T$. For a weakly interacting system with weak SU(2) symmetry breaking  $|g-g_{\uparrow\downarrow}|/g\ll1$, the main physics discussed above remains qualitatively unchanged,  except that the condensate momentum $k_0$ and spin polarization is slightly modified by the interaction \cite{YLi2012}. In this work, we focus on the two-minima region with $\Omega\in[0,4E_r]$, where the two degenerate ground states form the basis for the effective two-mode model in which spin squeezing can be realized as we will show now. We note that a similar two-mode model is considered in Refs.~\cite{engles1,engles2}.

\subsection{Two-Mode Model}
To this end, let us denote $a_L$ and $a_R$ as the annihilation operators for the ground states with momenta $-k_0$ and $k_0$, respectively. The projection operator for the ground state manifold with a fixed total number of particles $N$ is given by
\begin{equation}
\mathcal{P}_g = \sum_{N_{L}=0}^{N} \left|N_L,N_R\right\rangle\left\langle N_L, N_R\right|,
\label{A6}
\end{equation}
where $\left|N_L,N_R\right\rangle = (N_{L}! N_R!)^{-1/2} (a^\dagger_{L})^{N_{L}}(a^\dagger_{R})^{N_R}\left|0\right\rangle$ is the Fock state with $N_{L,R}$ atoms in each of the two degenerate modes under the constraint $N= N_{L} + N_{R}$. For dilute gas with weak interaction, the interaction Hamiltonian $H_\text{int}$ can be treated as a perturbation such that the main physics in the ground-state manifold can be efficiently depicted by the degenerate perturbation theory. Specifically, the first-order correction is given by
\begin{equation}
H_\text{int}^{(1)} = \mathcal{P}_g \mathcal{H}_\text{int} \mathcal{P}_g.
\label{A7}
\end{equation}
Since the projection operator $\mathcal{P}_g$ restrict all the calculations in the ground-state subspace, it is straightforward to expand the $H_\text{int}$ only by the two-mode field operators
 \begin{equation}
 \begin{aligned}
 	\boldsymbol{\hat\Psi} &= a_L \boldsymbol{\Phi}_{-k_0} + a_R \boldsymbol{\Phi}_{k_0} \\
 	&=
 	\frac{1}{\sqrt{L}}\begin{pmatrix}
 		\sin \theta e^{-ik_0 x} a_L +  \cos \theta e^{ik_0 x} a_R \\
 		- \cos \theta e^{-ik_0 x} a_L - \sin \theta e^{ik_0 x} a_R
 	\end{pmatrix},
\end{aligned}
\label{A8}
\end{equation}
using which we can straightforwardly obtain
\begin{equation}
H_\text{int}^{(1)} =-\chi F_z^2 + U'\hat{N}^2-U \hat{N}\,,
\label{A9}
\end{equation}
where $\hat{N} = \hat{N}_{L} + \hat{N}_{R}$ is the total number operator, and we have defined a set of spin field operators as
\begin{equation}
F_\mu =\frac{1}{2}\,  (a^\dag_L \; a^\dag_R)\, \sigma_\mu \left( \begin{array}{c} a_L \\ a_R \end{array} \right)\,,\;\;\mu=x,\, y,\, z.
\label{A9ad}
\end{equation}
In particular, $F_z = (\hat{N}_{L} - \hat{N}_{R})/2$ simply measures the population difference between the two modes. The coefficients in Eq.~(\ref{A9}) are given by
\begin{equation}
\begin{aligned}
	\chi &= \frac{(g+g_{\uparrow\downarrow})-(3g-g_{\uparrow\downarrow})\cos^2 2\theta }{2L},\\
	U' &= \frac{(g+g_{\uparrow\downarrow})(3-\cos^2 2\theta)}{8L},\\
	U &= \frac{(g+g_{\uparrow\downarrow}) + (g-g_{\uparrow\downarrow})\cos^2 2\theta}{4L}.
\end{aligned}
\label{A10}
\end{equation}
Since the total particle number $N$ is conserved, the last two terms in Eq.~(\ref{A9}) are constants that can be neglected. Consequently, within the first-order perturbation, the two-mode Hamiltonian is proportional to $F_z^2$ which is the canonical form of the one-axis-twisting (OAT) Hamiltonian~\cite{Kitagawa1993}.

 The strength of the nonlinearity spin interaction $\chi$ is a function of the interaction strengths ($g$ and $g_{\uparrow\downarrow}$) and $\theta$ which characterizes the Raman-induced SOC. In Fig.~\ref{Fig1}(b), we show the dependence of $\chi$ on $(g-g_{\uparrow\downarrow})$ and the Raman coupling strength $\Omega$. Specifically, the line at $\Omega=0$ ($\theta = 0$) denotes the case in absence of the SOC, and the system is then reduced to a conventional two-component (spin-1/2) BEC \cite{Ma2011,Sorensen2001}, with $\chi = -(g-g_{\uparrow\downarrow})/L$ only depending on the magnitude of the SU(2) symmetry breaking. Particularly, $\chi$ vanishes as $g=g_{\uparrow\downarrow}$. In sharp contrast, the presence of the SOC ($\Omega>0$) would lead to a finite $\chi$ even in the case of $g=g_{\uparrow\downarrow}$. Taking $g=g_{\uparrow\downarrow}$, we have
\begin{equation}
	\chi = \frac{g}{L}\sin^2 2\theta	= \frac{g}{L}\frac{\Omega^2}{16E_r^2}.
\label{A12}
\end{equation}
Note that since $g_{\uparrow\downarrow}$ can only take quite small difference with $g$ in the commonly used alkaline-metal atoms such as $^{87}$Rb and $^{23}$Na \cite{Zhai2015,Kawaguchi2012,Lin2011}, the SOC BEC provides a promising platform to realize a large nonlinearity on the order of density-density interaction $g$.

To follow the protocol in generating SSS under the OAT Hamiltonian $H_{\rm OAT} = -\chi F_z^2$, we need first to prepare the system in a coherent spin state (CSS) which is the eigenstate of the collective spin operator. We choose the spin operator to be $F_x= (a_L^\dagger a_R + a_R^\dagger a_L)/2$, and the corresponding CSS is taken to be the eigenstate with the maximum eigenvalue $N/2$, i.e. $\left|\text{CSS}\right\rangle = e^{-i\pi/2 F_y} \left|N,0\right\rangle$. Microscopically, this CSS state corresponds to a product state of identical $N$ particles with each particle in an equal-weight superposition of states $\boldsymbol{\Phi}_{\pm k_0}$, i.e.
\begin{equation}
\left|\text{CSS}\right\rangle = \prod_{i=1}^N\frac{\boldsymbol{\Phi}^i_{k_0} + \boldsymbol{\Phi}^i_{-k_0}}{\sqrt{2}}\,, \label{A13}
\end{equation}
which is referred to as a striped state \cite{Ho2011,YLi2012,YLi2013,engles2,Li2017} since the density profile of this state exhibits spatial oscillations. The CSS state (\ref{A13}) features equal quantum fluctuations in $F_y$ and $F_z$, i.e. $\Delta F_y^2 = \Delta F_z^2 = N/4$, with $\Delta F_{y,z}^2 = \left\langle F_{y,z}^2\right\rangle - \left\langle F_{y,z}\right\rangle^2$ and $N/4$ being the standard quantum limit (SQL). A state is called spin squeezed if its quantum variance along any direction in the $F_y$-$F_z$ plane is below the SQL. In general, the magnitude of squeezing is characterized by the squeezing parameter \cite{Kitagawa1993}
\begin{equation}
\xi^2 = \frac{4\Delta F_\text{min}^2}{N}, \label{A14}
\end{equation}
where $F_\text{min}$ corresponds to the spin along a direction in the $F_y$-$F_z$ plane with minimal spin variance. Hence, the SSS is featured by $\xi^2<1$.

To prepare the system in the $|{\rm CSS} \rangle$ state, we can add a strong effective transverse magnetic field along the $x$-axis which adds a term $J_0 F_x$ in the Hamiltonian. In practice, this can achieved by introducing an optical lattice potential $V_\text{ext} = V_0 \cos^2(k_0x) \sim V_0(e^{2ik_0x}+e^{-2ik_0x})/4$ \cite{note}. Viewed from the momentum space, the lattice potential would couple two states differing by a momentum of $\pm 2k_0$, hence can resonantly couple the two degenerate states $\boldsymbol{\Phi}_{\pm k_0}$. The lattice potential will also couple $\boldsymbol{\Phi}_{\pm k_0}$ to other states. However, such couplings are off-resonant. Projected onto the two-mode subspace, the effective two-mode Hamiltonian in the presence of the lattice potential reads
\begin{equation}
H_{\rm eff} = J_0 F_x-\chi F_z^2\,, \label{heff}
\end{equation}
where
\begin{equation}
J_0=-\frac{V_0\Omega}{16E_r}. \label{A26}
\end{equation}
We can then prepare the system in the ground state of Hamiltonian (\ref{heff}) in the limit of $|J_0| \gg N |\chi|$. Under this limit, the $F_x$ term dominates, and the ground state approximates the desired $|{\rm CSS} \rangle$ to a very good accuracy.

\subsection{Dynamical Generation of Spin Squeezing}
To dynamically generate the spin squeezing, we can simply quench the transverse magnetic field and follow the dynamics of the initial state $|{\rm CSS} \rangle$. We will consider two quench scenarios which we call Case 1 and Case 2.

{\em Case 1} --- In Case 1 quench protocol, the transverse magnetic field is completely turned off at $t=0$. Starting from $|{\rm CSS} \rangle$, the dynamics under the government of the OAT Hamiltonian has been well studied~\cite{Kitagawa1993}. It is shown that the squeezing parameter at time $t>0$ takes the following analytic form
\begin{equation}
\xi^2_\text{OAT}(t) = 1+\frac{1}{4}\left(N-1\right)\left(A-\sqrt{A^2+B^2}\right), \label{A15}
\end{equation}
with
\begin{equation}
\begin{aligned}
A &= 1-\cos^{N-2}\left(2\chi t\right), \\
B &= 2\sin \left(2\chi t\right) \cos^{N-1}\left(\chi t\right).
\end{aligned}
\label{A16}
\end{equation}
Under the condition $N|\chi|t<1$, $\xi^2_\text{OAT}$ takes its minimal value $\xi^2_\text{OAT} = (6N^2)^{-1/3}/2\sim N^{-2/3}$ at the optimal time $|\chi| t_\text{min} = 6^{1/6}N^{-2/3}$. For a given particle number $N$, a large nonlinearity can thus reduce the optimal squeezing time $t_\text{min}$ which is advantageous from an experimental point of view.

{\em Case 2} --- In Case 2 quench protocol, the strength of the transverse magnetic field is quenched to a smaller, but non-zero, value at $t=0$. For $t>0$, the system is governed by the Hamiltonian in the same form as Eq.~(\ref{heff}) except that $J_0$ is replaced by $J$ with $|J| < |J_0|$. This situation has been studied by Law \cite{Law2001}. For $|J| \ll |\chi|$, the squeezing behavior is similar to Case 1, as expected. For larger $|J|$, the remnant transverse magnetic field could play a significant role. Analytical results can only be obtained under the frozen spin approximation where it is assumed that $\left\langle F_x\right\rangle \approx N/2$ is fixed during the time evolution. This approximation is valid in the limit $|J| \gg |\chi|$. Under this approximation, the time evolution of the squeezing parameter is given by
\begin{widetext}
	\begin{equation}
	\xi^2_J = \cos^2(\epsilon t) + \frac{1}{2}\left(\frac{J^2}{\epsilon^2}+\frac{\epsilon^2}{J^2}\right)\sin^2 (\epsilon t) - \frac{1}{2}\left[
	\left(\frac{J^2}{\epsilon^2}-\frac{\epsilon^2}{J^2}\right)^2\sin^4 (\epsilon t) +
	\left(\frac{J}{\epsilon}-\frac{\epsilon}{J}\right)^2\sin^2 (2\epsilon t)
	\right]^{1/2}\,, \label{xiJ}
	\end{equation}
\end{widetext}
where $\epsilon \equiv \sqrt{J^2+N \chi |J|}$. Hence, $\xi_J$ exhibits periodic oscillation in time with period $\pi/\epsilon$.

\section{Numerical Simulation and Results}
The two-mode model provides a clear picture showing how spin squeezing can be realized. To confirm this, we now turn to numerical simulation under the full Hamiltonian (\ref{A1}). As the widely used mean-field treatment for atomic BEC ignores quantum fluctuations and correlations, in order to capture such quantum effects as spin squeezing, we have to go beyond the mean-field approach. To this end, we adopt the truncated Wigner method (TWM) \cite{Blakie2008,Altland2009} which takes into account the leading order of quantum fluctuation and is adequate for our purpose to investigate the spin squeezing property of the system. In the following, we first briefly describe the TWM, and then present our numerical results.

\subsection{Truncated Wigner Method}
In Wigner representation, a quantum state (or a density matrix) can be depicted by a quasi-probability distribution known as the Wigner function $W(\psi,\psi^*)$, where $\psi$ spans a coherent phase space satisfying $\hat{\psi}| \psi \rangle = \psi | \psi \rangle$. The expectation of an arbitrary operator $\hat{O}(\hat{\psi},\hat{\psi}^\dagger)$ can then be simply calculated through the average over the classical phase space, i.e.
\begin{equation}
\bar{O} = \int d^2\psi \,W(\psi,\psi^*) O_\text{cl}(\psi,\psi^*).
\label{A17}
\end{equation}
Here, $d^2\psi = d\psi d\psi^*$ denotes the complex integral and $O_\text{cl}(\psi,\psi^*)$ being the Weyl symbol of the operator $\hat{O}$, given explicitly by \cite{Altland2009}
\begin{equation}
O_\text{cl}=\int d^2\psi' \,\hat{O}(\hat{\psi} \rightarrow \psi-\psi'/2,\hat{\psi}^\dagger \rightarrow \psi^*+\psi'^*/2) e^{-|\psi'|^2/2}
\label{A18}
\end{equation}
as $\hat{O}$ is normally-ordered. For quantum dynamics, the Weyl symbol is time-dependent as the field operator $\hat{\psi}$ should satisfy the Heisenberg equation of motion $i\partial_t \hat{\psi} = [\hat{\psi},H]$. This operator equation, however, is in general too difficult to solve for a typical many-body system. The TWM is to assume that $\hat{\psi}$ follows a classical equation of motion $i\partial_t \psi = \delta \mathcal{H} /\delta \psi^*$ starting from an ensemble of initial states satisfying the quasi-probability distribution $W(\psi,\psi^*)$, and then the dynamic evolution of $\bar{O}$ can be simply obtained by the ensemble average. Here, $\mathcal{H}(\psi^*,\psi)$ denotes the energy functional of the system. Particularly in the context of atomic BEC, this classical equation of motion corresponds to the mean-field Gross-Pitaevskii (GP) equation.

To apply the TWM to the current system,  we first re-express the collective spin operators $F_{x,y,z}$ by the modes $a_\pm$ which is related with the $a_{L,R}$ by a unitary transformation $a_\pm = (a_L \pm a_R)/\sqrt{2}$. Then, the Wigner function of the CSS~(\ref{A13}) is given by \cite{Altland2009}
\begin{equation}
W(\xi_+,\xi_-) = \frac{2}{\pi^2} e^{-2|\xi_-|^2} \delta(|\xi_+|^2-N),
\label{A19}
\end{equation}
where $\xi_\pm$ being the c-number description of the modes $a_\pm$. On the other hand, the Weyl symbols of $F_{x,y,z}$ can be obtained in terms of $\xi_\pm$ from the integral Eq.~(\ref{A18}), and we have
\begin{equation}
\begin{aligned}
	F_{x,\text{cl}} & = \frac{1}{2}\left(|\xi_+|^2 - |\xi_-|^2\right) \equiv f_x,\\
	F_{y,\text{cl}} & = \frac{i}{2}\left(\xi_+^*\xi_- - \xi_-^*\xi_+\right) \equiv f_y,\\
	F_{z,\text{cl}} & = -\frac{1}{2}\left(\xi_+^*\xi_- + \xi_-^*\xi_+\right) \equiv f_z.
\end{aligned}
\label{A20}
\end{equation}
One can easily check $\bar{f}_{x} =N/2-1/4$, $\bar{f}_{y,z} = 0$ and $\Delta f_{y,z}^2 = \overline{f_{y,z}^2} - (\bar{f}_{y,z})^2 = N/4$ by substituting Eq.~(\ref{A19}) and (\ref{A20}) into Eq.~(\ref{A17}). Practically, it is  more convenient to rewrite $\xi_\pm$ by $\xi_\pm = \sqrt{N_\pm} e^{i\phi_\pm}$ and restrict $N_\pm$ by $N_+ + N_- =N$, which leads to the real distribution
\begin{equation}
W(\delta N,\delta \phi) = \frac{1}{2\pi}e^{-(N-\delta N)},
\label{A21}
\end{equation}
where $\delta N = N_+ - N_- \in [-N,N]$ is the number difference satisfying an exponential distribution with mean $N-1$ and variance $1$, whereas the relative phase $\delta \phi$ is totally random and uniformly distributed in the range $[-\pi,\pi)$. Obviously, the restriction affects nothing but shifts the mean spin $\bar{f}_{x}$ and the transverse variance $\Delta f_{y,z}^2$ by 1/4, which can be safely ignored as $N\gg1$.

Given system parameters $\Omega$, $g$ and $g_{\uparrow\downarrow}$, we implement the TWM as follows. We first prepare an ensemble of $10^3$ initial states
\begin{equation}
\begin{aligned}
	\mathbf{\Psi }_0 &= \xi_+ \boldsymbol{\Phi}_+ + \xi_- \boldsymbol{\Phi}_- \\
	&= \sqrt{\frac{N}{2}+\frac{\delta N}{2}}e^{i\delta\phi}\boldsymbol{\Phi}_+ + \sqrt{\frac{N}{2}-\frac{\delta N}{2}}\boldsymbol{\Phi}_-,
\end{aligned}
\label{A22}
\end{equation}
where $\boldsymbol{\Phi}_{\pm}  = \left( \boldsymbol{\Phi}_{k_0} \pm \boldsymbol{\Phi}_{-k_0} \right)/\sqrt{2}$, and $\delta N$ and $\delta \phi$ are sampled from the quasi-probability distribution Eq.~(\ref{A21}). We then let each of these sampled initial states evolve under the mean-field Hamiltonian by solving the coupled time-dependent GP equations
\begin{equation}
i{\partial \mathbf{\Psi }\left( x,t\right) }/{\partial t}=\left( H_{0}+%
\mathcal{G}\right) \mathbf{\Psi }\left( x,t\right),
\label{A23}
\end{equation}
where
$\mathcal{G}%
=\text{diag}\left( g\left\vert \psi _{\uparrow }\right\vert
^{2}+g_{\uparrow \downarrow }\left\vert \psi _{\downarrow }\right\vert
^{2},g\left\vert \psi _{\downarrow }\right\vert
^{2}+g_{\uparrow \downarrow }\left\vert \psi _{\uparrow }\right\vert
^{2}\right) $ characterizes the mean-field interactions. Finally, we project $\boldsymbol{\Psi}$ back to the ($\pm$)-basis to obtain
\begin{equation}
	\xi_\pm (t) = \int dx \, \boldsymbol{\Phi}^*_\pm(x,t) \boldsymbol{\Psi}(x,t) \,,
	\label{A24}
\end{equation}
using which the spin variance $\Delta f^2$ can be obtained by the ensemble average.

 Let us add a technical detail. Propagating up to thousands of GP equations over a long time period with low numerical error is indeed a challenging job. However, thanks to the development of the graphics processing unit (GPU)-based scientific computation, we designed a GPU solver that can propagate up to $10^3$ GP equations in a multi-thread manner \cite{GPU}. This allows us to carry out the calculation in a quite efficient way.

\subsection{Results}
\begin{figure}[t]
	\includegraphics[width=0.49\textwidth]{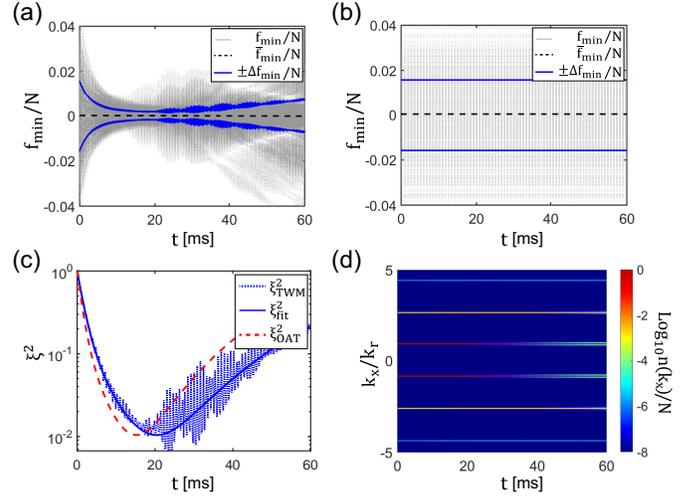}
	\caption{Squeezing dynamics for a BEC in boxed potential. (a) Spin evolution at  $\Omega = 2E_r$, $N=10^3$ and $gn\approx0.63E_r$ with $E_r = 5\times10^3(2\pi)$Hz, where dotted lines, dashed lines and solid lines denote the evolution of $f_\text{min}$, ensemble average of spin evolution $\bar{f}_\text{min}$, and the ensemble standard deviation $\Delta {f}_\text{min}$ along the direction with minimal ensemble variance. In the figure, $-\Delta {f}_\text{min}$ is also plotted such that the vertical distance between $\pm \Delta {f}_\text{min}$ can clearly demonstrate the spin squeezing. (b) Similar evolution with (a) but at $\Omega=0$, where no squeezing effect is observed. (c) Squeezing parameters as functions of t, where the dotted line, the dashed line and the solid line indicate the numerical result $\xi^2_\text{TWM}$, the analytical result $\xi^2_\text{OAT}$, and the fitting curve $\xi^2_\text{fit}$, respectively. (d) Logarithmic plot of the ensemble averaged momentum distribution $n(k_x)$.  }
	\label{Fig2}
\end{figure}

We now present our numerical results from the TWM and compare them with the analytical results based on the two-model model and the OAT Hamiltonian. We consider a homogeneous system
confined in a box with periodic boundary condition, and we only focus on the case with SU(2) symmetric interactions, i.e. $g=g_{\uparrow\downarrow}$, which is a rather accurate description for $^{87}$Rb and $^{23}$Na \cite{Kawaguchi2012,Lin2011}.

{\em Case 1} --- We first consider the Case 1 quench where at $t=0$ the lattice potential is suddenly turned off, hence in solving the GP equations, the external potential is taken to be $V_{\rm ext}=0$. The gray dotted lines in Fig.~\ref{Fig2}(a) show the spin evolution $f_\text{min}(t)$, starting from the $10^3$ sampled initial states using Eq.~(\ref{A22}), where $f_\text{min}(t)$ denotes the polarization along the direction in the $f_y$-$f_z$ plane in which the variance takes the minimal value. In the calculation, we consider $N=10^3$ $^{87}$Rb atoms, whose $s$-wave scattering length is $a_s = 101.8 a_B$ with $a_B$ being the Bohr radius, in a rectangle box with geometry $L=L_x=10\mu m$ and $L_y=L_z=0.5\mu m$, and take $\Omega = 2 E_r$ with $E_r = 5\times10^3(2\pi)$Hz, such that the interaction $g$ satisfy $gn \approx 0.63 E_r $ with $n=4\times10^{14}\text{cm}^{-3}$ being the averaged atomic density. The ensemble average over the whole sample is $\bar{f}_\text{min}(t)= 0$ as indicated by the black dashed line, and the standard deviation of the whole sample $\Delta f_\text{min}(t)$ is shown as the blue solid line. Particularly at the initial time $t=0$, $\Delta f_\text{min}(0) = \sqrt{N}/2$ represents the SQL. It can be seen that, at later time, $\Delta f_\text{min}(t)$ drops below the SQL, and reaches a minimum before rising again. This is a clear manifestation of spin squeezing. As a comparision, we also show the dynamical evolution in the case without SOC (by setting $\Omega = 0$) in Fig.~\ref{Fig2}(b), and apparently no squeezing effect is observed as $\Delta f_\text{min}(t)$ remains fixed at the SQL, which is in agreement with our analytical discussion.

\begin{figure}[t]
	\includegraphics[width=0.466\textwidth]{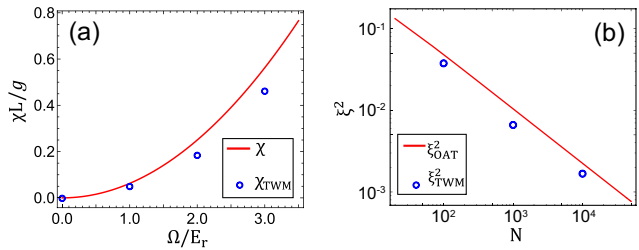}
	\caption{(a) Analytical $\chi_\text{OAT}$ and the numerically extracted $\chi_\text{TWM}$ as functions of $\Omega$, where $N=10^3$ and $gn \approx 0.63 E_r$ are fixed. (b) Dependence of the optimal squeezing parameter on $N$, where solid line and hollow circles correspond to the analytical result and the numerical result, respectively. In the calculation, we fix $\Omega = 2 E_r$.  }
	\label{Fig3}
\end{figure}

In Fig.~\ref{Fig2}(c), we show the squeezing parameter $\xi^2_\text{TWM}$ obtained from the TWM (blue dotted line) corresponding to the dynamics shown in Fig.~\ref{Fig2}(a). In comparison, the analytical result $\xi^2_\text{OAT}$ obtained from Eq.~(\ref{A15}) (red dashed line) is also plotted. The two results exhibit good qualitative agreement. The quantitative discrepancy between $\xi^2_\text{TWM}$ and $\xi^2_\text{OAT}$ mainly lies in the fact that the $\xi^2_\text{TWM}$ has oscillations as $t$ increases, and that the optimal squeezing time $t_\text{min}$ of $\xi^2_\text{TWM}$ is slightly larger than that of $\xi^2_\text{OAT}$. These descrepancies can be mainly attributed to finite-size effect and high-order scattering that are not captured by the effective two-mode model. The former arises from the fact that the plane wave function $\boldsymbol{\Phi}_{\pm k_0}$ in a homogeneous system of length $L$ with periodic boundary condition is actually a superposition of plane waves with momenta $c k_0$, with $c$ being odd integers and $c=\pm1$ being the dominant modes \cite{YLi2013}; and the latter lies in the fact that a small portion of the particles are pumped away to the excited states near $\pm k_0$. To see this, we plot the logarithmic of the ensemble averaged momentum distribution $n(k_x) = |\int dx \boldsymbol{\Psi}(x) e^{-i k_x x}|^2$ as a function of $t$ in Fig.~\ref{Fig2}(d) where the two effects are clearly demonstrated. The particles out of the ground-state subspace mainly experience a smaller nonlinear spin interaction strength $\chi$ due to a small $\theta$ (see Eq.~(\ref{A12})), which effectively reduces the overall nonlinearity and leads to longer $t_\text{min}$. We can obtain the effective spin nonlinearity from the numerical results by fitting the $\xi^2_\text{TWM}$ with the analytical result $\xi^2_\text{OAT}$ and treating $\chi$ as a fitting parameter. The fitting curve is shown as the blue solid line in Fig.~\ref{Fig2}(c), from which we can extract the fitted effective nonlinear spin interaction strength $\chi_\text{TWM} \approx 0.19g/L$ which is slightly smaller than $\chi=0.25g/L$ as calculated from the two-mode model using Eq.~(\ref{A12}).

We perform similar calculations for other values of $\Omega$, and show the corresponding $\chi_\text{TWM}$ as circles in Fig.~\ref{Fig3}(a). Moreover, the analytical $\chi$ (Eq.~(\ref{A12})) is shown as the solid line as a comparison. It turns out that the numerical nonlinearity $\chi_\text{TWM}$ is approximately proportion to $\Omega^2$, as $\chi$. In addition, we also carry out the similar calculation for different values of $N=10^2$, $10^3$ and $10^4$ at fixed $\Omega=2E_r$, and show the maximal squeezing parameter $\xi^2_\text{TWM}$ as a function of $N$ by circles in Fig.~\ref{Fig3}(b). Again, the analytical predication is also shown by the solid line. The result confirms the particle number scaling $\xi^2_\text{TWM}\sim N^{-2/3}$.

\begin{figure}[t]
	\includegraphics[width=0.49\textwidth]{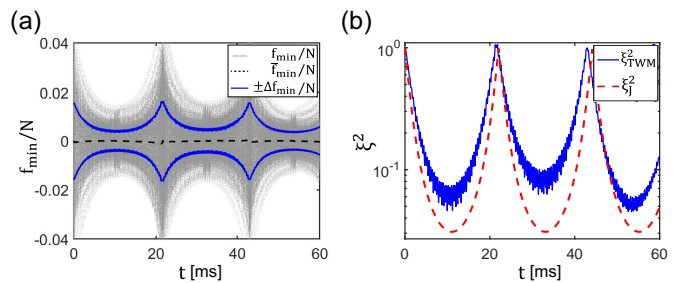}
	\caption{Squeezing dynamics of a BEC in the presence of an optical lattice. (a) Spin evolution with dotted lines, dashed lines and solid lines denoting the evolution of $f_\text{min}$, ensemble average of spin evolution $\bar{f}_\text{min}$ and the ensemble standard deviation $\Delta {f}_\text{min}$ along the direction with minimal ensemble variance. (b) Evolution of the squeezing parameters obtained by the TWM $\xi^2_\text{TWM}$ (solid line) and from the effective model $\xi^2_\text{J}$ (dashed line) Eq.~(\ref{heff}). In our calculation, we take $V_0 = 0.04E_r$, $\Omega = 2E_r$, $N = 10^3$, $gn \approx 0.63 E_r$ with $E_r = 5\times10^3(2\pi)$Hz, and the resulting tunneling strength and the nonlinearity satisfy $|J|/\chi \approx 31.8$. }
	\label{Fig4}
\end{figure}

{\em Case 2} --- We now consider the Case 2 quench in which the initial strong lattice potential is weakened but not turned off. In solving the GP equation, we thus take $V_{\rm ext} = V_0 \cos^2 (k_0 x)$ with $V_0 = 0.04E_r$, and fix other parameters being the same with those used in the Case 1 quench. As a result, the corresponding two-mode tunneling is $J = -5\times 10 ^{-3}E_r$ and satisfies the condition $|J|/\chi \approx 31.8 \gg 1$ required by the frozen spin approximation. The numerical results are presented in Fig.~\ref{Fig4}, in which the evolutions of $f_{\rm min}$ and its variance $\Delta f_{\rm min}$ are plotted in Fig.~\ref{Fig4}(a), and the corresponding squeezing parameter $\xi^2$ is presented in Fig.~\ref{Fig4}(b), where we also plot the analytical result from the two-mode calculation $\xi^2_J$ as expressed in Eq.~(\ref{xiJ}).  One can see that the numerically obtained squeezing parameter and the analytical result are in good qualitative agreement. Again, the fast oscillations in the numerical curve may be attributed to the multi-mode effects that cannot be captured by the effective two-mode model.

\subsection{Discussion}
In the above, we have considered a homogeneous system. In practice, most BEC experiments are done in the presence of harmonic trapping potentials. The presence of a trapping potential, however, would not change the main physics. For such a trapped condensate subjected to Raman-induced SOC, we still have two degenerate ground states when the Raman coupling strength is below a threshold, and the wavefunction of the two degenerate ground states are still approximately given by Eq.~(\ref{A4}) with an additional Gaussian envelop function~\cite{zhu,YLi2012}. Hence the results obtained above remains qualitatively valid.

Finally, let us discuss how to detect the spin squeezing by measuring spin variances. The measurement of the spin variance in $F_z$ is straightforward, since the two modes $L$ and $R$ are completely separable in the momentum space and will be spatially separated after a time of flight, and $F_z$ is just the population difference between the two modes. For the measurement of an arbitrary spin in the $F_y$-$F_z$ plane, the so-called spin-noise tomography \cite{Riedel2010,Gross2010} can be implemented. Practically, one can pulse on an evolution under the government of $H_0 = J_0 F_x$ for a proper duration to rotate the spin onto the measurable $z$-axis, and then the population difference detection can be carried out directly.

\section{Conclusion}
To conclude, we have demonstrated the realization of spin squeezing in a spin-orbit coupled atomic condensate. We show that the SOC provides a crucial new control knob to facilitate spin squeezing, circumventing the restriction on interaction strengths in the absence of the SOC. We have constructed a simple two-mode model to explain the main physics, which is confirmed by a detailed numerical simulation beyond the usual mean-field treatment.

We would also like to point out that, though the above discussions are performed on a spin-1/2 BEC with Raman induced SOC, the basic concept behind is not limited to this particular scenario. For example, it can be generalized to a system with spin-and-orbit-angular-momentum coupling \cite{DeMarco2015,Sun2015,Qu2015,Hu2015,Chen2016,Chen2019}, which has been experimentally realized very recently \cite{HChen2018,PChen2018,Zhang2018}, where a pair of lower-lying mode with opposite orbital angular momenta can play the role of modes L and R in our system. Additionally, similar discussion can be performed in SOC systems with higher-spins \cite{Lan2014,Natu2015,Sun2016,Yu2016,Martone2016,Campbell2016}, in which richer squeezing effects (e.g. spin-nematic squeezing \cite{Hamley2012}) may be observed. Our work adds yet another interesting piece of physics that synthetic spin-orbit coupling brings to the field of cold atoms.

\begin{acknowledgments}
L. Chen would like to thank X. Peng for helpful discussion, and M. Chang for sharing the GPU during the calculation. LC acknowledge supports from the NSF of China (Grant No. 11804205), and the Beijing Outstanding Young Scientist Program hold by H. Zhai; YZ acknowledge supports from the NSF of China (Grant No. 11674201); HP acknowledges supports from the US NSF and the Welch Foundation (Grant No. C-1669).
\end{acknowledgments}


\begin{thebibliography}{99}
\bibitem{Kitagawa1993} M. Kitagawa and M. Ueda, Phys. Rev. A \textbf{47}, 5138 (1993).

\bibitem{Ma2011} J. Ma, X. Wang, C. P. Sun, and F. Nori, Phys. Rep. \textbf{509}, 89 (2011).

\bibitem{Pezze2018}  L. Pezz{\`e}, A. Smerzi, M. K. Oberthaler, R. Schmied, and P. Treutlein, Rev. Mod. Phys. \textbf{90}, 035005 (2018).

\bibitem{Wineland} D. J. Wineland, J. J. Bollinger, W. M. Itano, F. L. Moore, and D. J. Heinzen, Phys. Rev. A \textbf{46}, R6797 (1992); D. J. Wineland, J. J. Bollinger, W. M. Itano, and D. J. Heinzen, ibid. \textbf{50}, 67 (1994); V. Meyer, M. Rowe, D. Kielpinski, C. Sackett, W. Itano, C. Monroe, and D. Wineland, Phys. Rev. Lett. \textbf{86}, 5870 (2001).

\bibitem{Cronin2009} A. D. Cronin, J. Schmiedmayer, and D. E. Pritchard, Rev. Mod. Phys. \textbf{81}, 1051 (2009).

\bibitem{Bigelow2001} N. Bigelow, Nature \textbf{409}, 27 (2001); E. S. Polzik, ibid. \textbf{453}, 45 (2008).

\bibitem{Gravity} D. F. Walls and P. Zoller, Phys. Lett. A \textbf{85}, 118 (1981); K. Goda, O. Miyakawa, E. E. Mikhailov, S. Saraf, R. Adhikari, K. McKenzie, R. Ward, S. Vass, A. J. Weinstein, and N. Mavalvala, Nature Phys. \textbf{4}, 472 (2008).

\bibitem{Sorensen2001} A. S{\o}rensen, L.-M. Duan, J. I. Cirac, and P. Zoller, Nature (London) \textbf{409}, 63 (2001).

\bibitem{Esteve2008} J. Est{\`e}ve, C. Gross, A. Weller, S. Giovanazzi, and M. K. Oberthaler, Nature (London) \textbf{455}, 1216 (2008).

\bibitem{Riedel2010} M. F. Riedel, P. B{\"o}hi, Y. Li, T. W. H{\"a}nsch, A. Sinatra, and P. Treutlein, Nature (London) \textbf{464}, 1170 (2010).

\bibitem{Gross2010} C. Gross, T. Zibold, E. Nicklas, J. Est{\`e}ve, and M. K. Oberthaler, Nature (London) \textbf{464}, 1165 (2010).

\bibitem{Hamley2012} C. D. Hamley, C. S. Gerving, T. M. Hoang, E. M. Bookjans, and M. S. Chapman, Nature Phys. \textbf{8}, 305 (2012).

\bibitem{Kawaguchi2012} K. Kawaguchi, and M. Ueda, Phys. Rep. \textbf{520}, 253 (2012).

\bibitem{Galitski2013} V. Galitski and I. B. Spielman, Nature \textbf{494}, 49 (2013).

\bibitem{Goldman2014} N. Goldman, G. Juzeli\={u}nas, P. \"{O}hberg, and I.
  B. Spielman, Rep. Prog. Phys. \textbf{77}, 126401 (2014).

\bibitem{Zhai2015} H. Zhai, Rep. Prog. Phys. \textbf{78}, 026001 (2015).

\bibitem{WZhang2018} \textit{Synthetic Spin-Orbit Coupling in Cold Atoms}, edited by W. Zhang, W. Yi, and C. A. R. Sá Melo (World Scientific, Singapore, 2018).

\bibitem{Lin2011} Y.-J. Lin, K. Jim\'{e}nez-Garc\'{\i}a, and I. B. Spielman, Nature (London). \textbf{471}, 83 (2011).

\bibitem{YLi2012} Y. Li, L. P. Pitaevskii, and S. Stringari, Phys. Rev. Lett. \textbf{108}, 225301 (2012).

\bibitem{engles1} J. Hou, X.-W. Luo, K. Sun, T. Bersano, V. Gokhroo, S. Mossman, P. Engels, and C. Zhang, Phys. Rev. Lett. {\bf 120}, 120401 (2018).

\bibitem{engles2} T. M. Bersano, J. Hou, S. Mossman, V. Gokhroo, X.-W. Luo, K. Sun, C. Zhang, and P. Engels, Phys. Rev. A {\bf 99}, 051602(R) (2019).

\bibitem{Ho2011} T.-L. Ho and S. Zhang, Phys. Rev. Lett. \textbf{107}, 150403 (2011).

\bibitem{YLi2013} Y. Li, G. I. Martone, L. P. Pitaevskii and S. Stringari, Phys. Rev. Lett. \textbf{110}, 235302 (2013).

\bibitem{Li2017} J.-R. Li, J. Lee, W. Huang, S. Burchesky, B. Shteynas, F. Cagri Top, A. O. Jamison, and W. Ketterle, Nature \textbf{543}, 91 (2017).

\bibitem{note}Alternatively, this can be achieved by adding an additional pair of Raman beams as is done in the experiments reported in Refs.~\cite{engles1,engles2}.

\bibitem{Law2001} C. Law, H. Ng, and P. Leung, Phys. Rev. A \textbf{63}, 055601 (2001).

\bibitem{Blakie2008} P. B. Blakie, A. S. Bradley, M. J. Davis, R. J. Ballagh, and C. W. Gardiner, Adv. Phys. \textbf{57}, 363 (2008).

\bibitem{Altland2009} A. Altland, V. Gurarie, T. Kriecherbauer, and A. Polkovnikov, Rev. A \textbf{79}, 042703 (2009).

\bibitem{GPU} In our numerical calculation, we use a Nvidia Titan V with double-precision-floating-point performance up to 6.9 Teraflops, which consequently provides a computational acceleration ratio $\sim 100$ comparing with a single-thread GP solver working on CPU with clock speed of 2.2GHz.

\bibitem{zhu}C. Zhu, L. Dong, and H. Pu, J. Phys. B {\bf 49}, 145301 (2016).

\bibitem{DeMarco2015} M. DeMarco and H. Pu, Phys. Rev. A. \textbf{91}, 033630
(2015).

\bibitem{Sun2015} K. Sun, C. Qu, and C. Zhang, Phys. Rev. A. \textbf{91}, 063627
(2015).

\bibitem{Qu2015} C. Qu, K. Sun, and C. Zhang, Phys. Rev. A. \textbf{91}, 053630
(2015).

\bibitem{Hu2015} Y.-X. Hu, C. Miniatura, and B. Gr\'{e}maud, Phys. Rev. A. \textbf{92}, 033615
(2015).

\bibitem{Chen2016} L. Chen, H. Pu, and Y. Zhang, Phys. Rev. A \textbf{93}, 013629 (2016).

\bibitem{Chen2019} X.-L. Chen, S.-G. Peng, P. Zou, X.-J. Liu, H. Hu, arXiv:1901.02595.

\bibitem{HChen2018} H.-R. Chen, K.-Y. Lin, P.-K. Chen, N.-C. Chiu,   J.-B. Wang, C.-A. Chen, P.-P. Huang, S.-K. Yip, Y.  Kawaguchi, and Y.-J. Lin, Phys. Rev. Lett. \textbf{121}, 113204 (2018).

\bibitem{PChen2018} P.-K. Chen, L.-R. Liu, M.-J. Tsai, N.-C. Chiu, Y. Kawaguchi, S.-K. Yip, M.-S. Chang, and Y.-J. Lin, Phys. Rev. Lett. \textbf{121}, 250401 (2018).

\bibitem{Zhang2018} D. Zhang, T. Gao, P. Zou, L. Kong, R. Li, X. Shen, X.-L. Chen, S.-G. Peng, M. Zhan, H. Pu, and J. Kaijun, Phys. Rev. Lett. \textbf{122}, 110402 (2019).

\bibitem{Lan2014} Z. H. Lan, and P. \"{O}hberg, Phys. Rev. A \textbf{89}, 023630 (2014).

\bibitem{Natu2015} S. S. Natu, X. P. Li, and W. S. Cole, Phys. Rev. A \textbf{91}, 023608 (2015).

\bibitem{Campbell2016} D. L. Campbell, R. M. Price, A. Putra, A. Vald\'{e}s-Curiel, D. Trypogeorgos, and I. B. Spielman, Nat. Commun. \textbf{7}, 10897 (2016).

\bibitem{Sun2016} K. Sun, C. Qu, Y. Xu, Y. Zhang, and C. Zhang, Phys. Rev. A \textbf{93}, 023615 (2016).

\bibitem{Yu2016} Z.-Q. Yu, Phys. Rev. A \textbf{93}, 033648 (2016).

\bibitem{Martone2016} G. I. Martone, F. V. Pepe, P. Facchi, S. Pascazio, and S. Stringari, Phys. Rev. Lett. \textbf{117}, 125301 (2016).
\end{thebibliography}
\end{document}